\documentclass[twocolumn, a4paper, superscriptaddress, notitlepage, floatfix, aps, prl]{revtex4-1}
\pdfoutput=1
\usepackage[utf8]{inputenc}
\usepackage[usenames]{color}
\usepackage{graphicx}
\usepackage{amsmath, amssymb, amsfonts, mathtext, array}
\usepackage{hhline,layout}
\usepackage{titlesec}
\usepackage[pdftex,unicode,colorlinks,citecolor=blue,linkcolor=blue,]{hyperref}

\makeatletter
\renewcommand\thefigure{\@arabic\c@figure}
\renewcommand\fnum@figure{\figurename~\thefigure}
\newcommand\onlinecap{\renewcommand\fnum@figure{\figurename~\thefigure~(color online)}}
\makeatother

\begin{document}

\title{Beating the standard sensitivity-bandwidth limit of cavity-enhanced interferometers with internal squeezed-light generation}

\author{M. Korobko}
\affiliation{Institut f\"ur Laserphysik und Zentrum f\"ur Optische Quantentechnologien,
Universit\"at Hamburg, Luruper Chaussee 149, 22761 Hamburg, Germany}
\author{L. Kleybolte}
\affiliation{Institut f\"ur Laserphysik und Zentrum f\"ur Optische Quantentechnologien,
Universit\"at Hamburg, Luruper Chaussee 149, 22761 Hamburg, Germany}
\author{S. Ast}
\affiliation{Institut f\"ur Gravitationsphysik, Leibniz Universit\"at Hannover and Max-Planck-Institut f\"ur
Gravitationsphysik (Albert-Einstein-Institut), Callinstra{\ss}e 38, 30167 Hannover, Germany}
\author{H. Miao}
\affiliation{Institute of Gravitational Wave Astronomy, University of Birmingham, Birmingham B15 2TT, United Kingdom }
\author{Y. Chen}
\affiliation{Caltech CaRT, Pasadena, California 91125, USA}
\author{R. Schnabel}
\affiliation{Institut f\"ur Laserphysik und Zentrum f\"ur Optische Quantentechnologien,
Universit\"at Hamburg, Luruper Chaussee 149, 22761 Hamburg, Germany}

\begin{abstract}
The shot-noise limited peak sensitivity of cavity-enhanced interferometric measurement devices, such as gravitational-wave detectors, can be improved by increasing the cavity finesse, even when comparing fixed intra-cavity light powers. 
For a fixed light power inside the detector, this comes at the price of a proportional reduction in the detection bandwidth. 
High sensitivity over a large span of signal frequencies, however, is essential for astronomical observations. 
It is possible to overcome this standard sensitivity-bandwidth limit using non-classical correlations in the light field. 
Here, we investigate the \textit{internal squeezing} approach, where the parametric amplification process creates a non-classical correlation directly inside the interferometer cavity. 
We analyse the limits of the approach theoretically, and measure 36\% increase in the sensitivity-bandwidth product compared to the classical case. 
To our knowledge this is the first experimental demonstration of an improvement in the sensitivity-bandwidth product using internal squeezing, opening the way for a new class of optomechanical force sensing devices.
 \end{abstract}
 
 \maketitle
 
\noindent\textit{Introduction.---} 
Optical cavities can be used to enhance the sensitivity of interferometric measurements of small signals caused by a weak classical force acting on a movable mirror. 
The motion of the mirror produces a phase modulation on the light field, which then gets enhanced by constructive interference with itself on the cavity round trip.
For any given light power inside the detector cavity, increasing the cavity finesse improves the shot-noise limited sensitivity, but is necessarily accompanied by a proportional reduction of the detection bandwidth \cite{Caves1982, Clerk2010}.
This effect limits the performance of all gravitational-wave detectors (Advanced LIGO, GEO600, Advanced Virgo, KAGRA) \cite{Abadie2011, Aso2013, Aasi2015, Acernese2015}. 
Typical gravitational-wave signals require high but also broadband sensitivity: the signal from a binary black hole merger, such as the one detected in September 2015 \cite{Abbott2016}, sweeps through the frequencies of the interferometer's detection band.

According to the Heisenberg uncertainty principle, one has to increase the uncertainty in the light's amplitude quadrature in order to improve the measurement sensitivity by decreasing the uncertainty in the light's phase quadrature.
Since energy is needed to increase the uncertainty, the sensitivity limit of an interferometer is set by the optical energy inside the cavity \cite{92BookBrKh, 00p1BrGoKhTh}.
In a more general case of arbitrary signal waveforms this consideration leads to the Quantum Cramer-Rao Bound (QCRB) for the estimation of signal in gaussian quantum noise: at each signal frequency the maximal phase sensitivity is set by the size of the amplitude quadrature uncertainty at the same frequency \cite{Tsang2011, Miao2016}. 

Based on the QCRB, first of all, the concept of enhancing the sensitivity with optical cavity can be understood. 
Both amplitude and phase quadratures resonate inside the cavity, and have their uncertainties amplified within the bandwidth of the resonance, and attenuated at other frequencies. 
In the case of a coherent input field and a simple Fabry-Perot cavity the state remains coherent inside the cavity.
The standard sensitivity-bandwidth limit is defined as the maximum product of a peak sensitivity $\mathcal{S}$ and a detection bandwidth $\mathcal{B}$, that can be achieved using coherent states of light and a given light power $P_c$ inside the cavity \cite{Mizuno1995a}: $\mathcal{S}\times\mathcal{B} \leq 8 \pi P_c /(\hbar \lambda L)$, where $\lambda$ is the optical wavelength,  $L$ is the cavity length and $\hbar$ is the reduced Plank constant.

We introduce a set of different strategies for improving the sensitivity of a cavity-enhanced interferometer beyond the standard sensitivity-bandwidth limit. 
The first approach is called the {\it white-light cavity} effect. 
It broadens the cavity resonance without changing the finesse, in which case the uncertainty of the amplitude quadrature must increase above the vacuum level.
It was proposed recently that the white-light cavity effect can be achieved by using an anomalously dispersive medium inside the interferometer \cite{Wicht1997, Wicht2000, Wise2004a, Wise2005, Pati2007, Yum2013, Zhou2015, Miao2105, Qin2015}.

The second approach is called {\it external squeezing}. 
In this case, the uncertainty of the optical field injected in the interferometer is squeezed below the vacuum level in the phase quadrature, without influencing the signal enhancement due to the optical cavity \cite{Caves1980a, Caves1981, Schnabel2010, Abadie2011, Abadie2013}.
The bandwidth remains unchanged, and hence the standard sensitivity-bandwidth limit is surpassed due to the increased peak signal-to-noise ratio.

The third approach is {\it internal squeezing}.
Here, a squeezed state of light is produced inside the detector's Fabry-Perot cavity, for instance using an optical parametric amplifier \cite{Rehbein2005, V.Peano2015, Somiya2016}.
In contrast to external squeezing, in this approach the phase quadrature squeezing happens mainly inside the optical cavity linewidth and affects both the noise and the signal. 
The amplitude quadrature uncertainty is correspondingly increased above the vacuum level, and in accordance with the QCRB the sensitivity increases: the noise is squeezed more than the signal is deamplified.
The detection bandwidth narrows in this case, but the peak sensitivity is increased even more strongly, which allows the standard sensitivity-bandwidth limit to be surpassed. 

In this work we analyse the third approach theoretically and report on a proof-of-principle experiment in which the standard sensitivity-bandwidth limit was surpassed by 36\%.
We note that our work does not consider quantum radiation pressure noise, effectively assuming an infinite mass of the test mirrors.
In practice, radiation pressure noise can generally be eliminated by increasing the mass of the mirrors or by using Quantum Non-Demolition measurement techniques \cite{92BookBrKh, Kimble2000, Khalili2014, Miao2014, Aasi2015}.

\begin{figure}
	\includegraphics{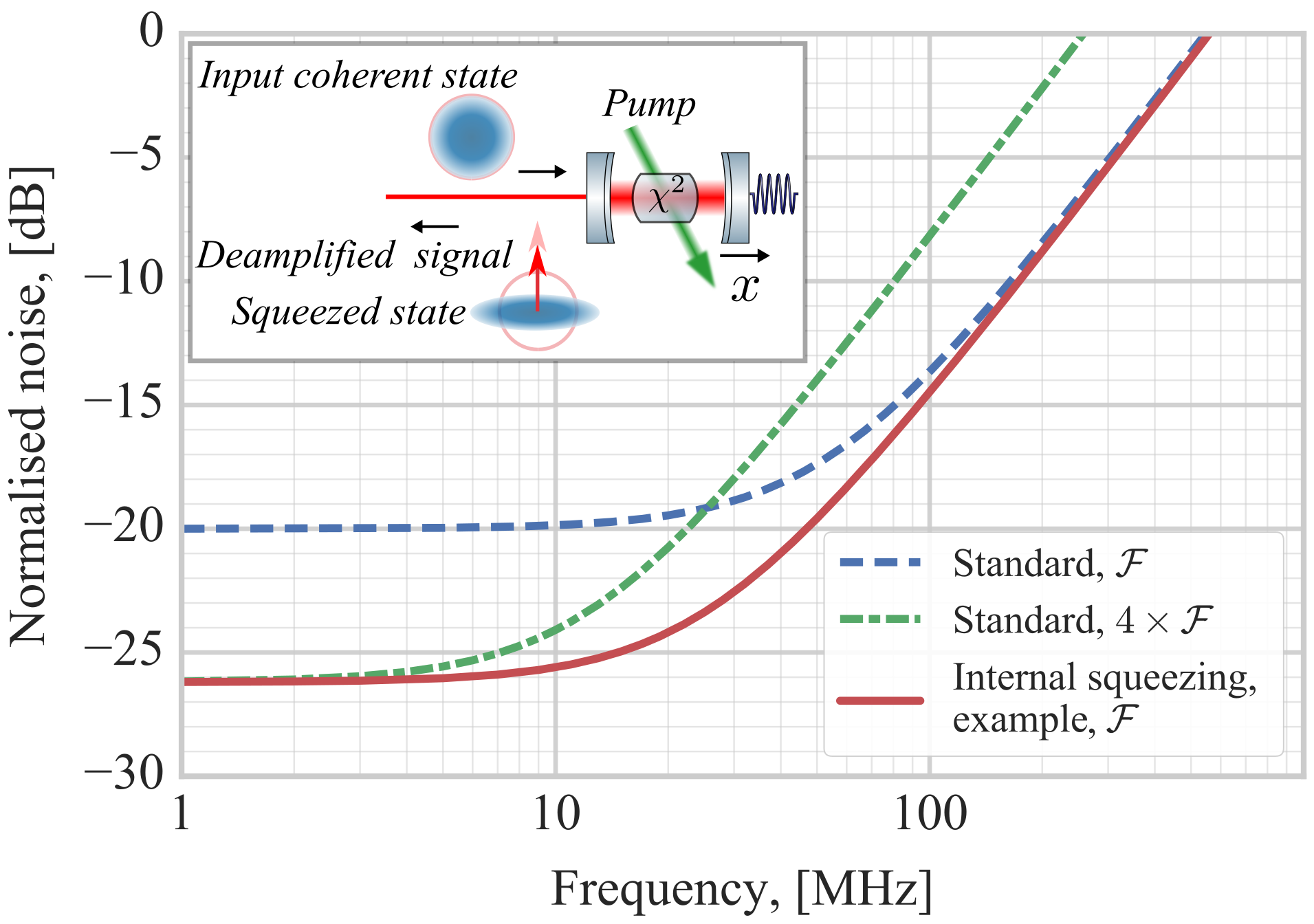}
	\onlinecap\caption{The three curves show the quantum measurement noise of a cavity-enhanced interferometer with the same coherent light power in its arms, normalised to a phase signal optical transfer function. The peak sensitivity $\mathcal{S}$ is defined as inverse of the minimum of the curves; the bandwidth $\mathcal{B}$ is the frequency at which the noise rises by 3\,dB above its minimal value. The standard sensitivity-bandwidth product remains constant for a given coherent light power inside the cavity: to increase the peak sensitivity by 6\,dB, the finesse $\mathcal{F}$ has to be increased by a factor of 4, thus the bandwidth decreases by the same amount (compare blue dashed and green dashed-dotted curves). The internal squeezing approach is depicted schematically on the subplot: a phase quadrature of the input coherent light field is squeezed inside the cavity, deamplifying the signal at the same time. Signal deamplification is less than the amount of squeezing on the detector, so the sensitivity increases. Increased amplitude quadrature, according to the QCRB, leads to enhancement of the sensitivity-bandwidth product beyond the standard limit. Therefore when increasing the peak sensitivity by 6\,dB of internal squeezing the resulting bandwidth is broader (red solid curve) than the one achievable classically.}\label{fig:int_vs_ext}
 \end{figure}

\noindent\textit{General concept.---}
 We consider the propagation of a signal through a Fabry-Perot cavity with a nonlinear crystal inside, see Fig.\,\ref{fig:int_vs_ext}.
 Pumping the crystal with light of the doubled frequency leads to optical parametric amplification of the cavity mode. 
 The highest squeeze factor inside the cavity is achieved around cavity resonance and is limited to 6 dB.
At this level the threshold for optical parametric oscillation is reached, and the amplified amplitude quadrature becomes unstable and causes lasing \cite{Milburn1981,Collett1984}. 
 However, the squeeze factor outside the cavity is not fundamentally limited, due to destructive interference between the incoming coherent field and outgoing squeezing \cite{Schnabel2016}. 
 On the other hand, the signal originates from the inside of the cavity, and does not experience this such interference.
 Therefore the deamplification in the signal remains limited to 6\,dB.
 The resulting difference between noise squeezing and signal deamplification constitutes the gain in the signal-to-noise ratio (SNR), which represents the sensitivity of the detector.   
 On the other hand, the bandwidth gets reduced, as the internal squeezing increases the sensitivity only inside the cavity linewidth, and leaves it unchanged outside. 
 Despite this, the sensitivity-bandwidth product is enhanced, according to the QCRB, as we amplify the amplitude quadrature fluctuations inside the cavity.

We present a simplified treatment of a mathematical model of the system, leaving a rigorous treatment for the supplementary material.
We define three quantities that influence the cavity bandwidth: cavity decay rate through the coupling mirror, squeezing rate and the roundtrip optical loss rate, correspondingly
\begin{equation}
\gamma_c = \frac{c t_{\rm c}^2}{4 L}, \quad \gamma_s = \frac{q c}{L}, \quad \gamma_l = \frac{c l^2}{4 L},
\end{equation}
where $c$ is the speed of light, $L$ is the optical length of the cavity, $t_{\rm c}$ is the amplitude transmissivity of the coupling mirror, $q$ is the squeeze factor on a single pass through a crystal,  $l^2$ is the round-trip internal loss without the transmission of the coupling mirror.

From the optical fields' input-output relations we derive the power spectral density of noise of the output field detected by a balanced homodyne detector
\begin{equation}\label{eq:noise_tf}
S_n (\Omega) = 1 - \frac{4 \gamma_c \gamma_s \eta}{(\gamma_c + \gamma_l + \gamma_s)^2 + \Omega^2},
\end{equation}
where $\Omega$ is the signal frequency, $\eta$ is total detection efficiency including light propagation and the quantum efficiency of the homodyne. 
Correspondingly, for the optical transfer function $T(\Omega)$ of the phase modulation signal through the cavity to the detector we find
\begin{equation}\label{eq:signal_tf}
|T(\Omega)|^2\approx  \frac{8 \pi P_c}{\hbar \lambda L}\frac{\gamma_c \eta}{(\gamma_c + \gamma_l + \gamma_s)^2 +  \Omega^2}.
\end{equation}

The equations above lead to the definition of the common bandwidth for the noise and the signal transfer functions $\Gamma = \gamma_c + \gamma_l + \gamma_s.$
Then we define the sensitivity
\begin{equation}
\frac{|T(\Omega)|^2}{S_n(\Omega)} = \frac{8 \pi P_c}{\hbar \lambda L}\frac{\gamma_c \eta}{\Gamma^2 - 4 \gamma_c \gamma_s \eta+ \Omega^2}.
\end{equation}
It's  peak value $\mathcal{S} \equiv |T(0)|^2/S_n(0)$  and bandwidth $\mathcal{B}$ are given by
\begin{align}\label{eq:snr}
\mathcal{S} &= \frac{8 \pi P_c}{\hbar \lambda L}\frac{\gamma_c \eta}{\mathcal{B}^2}, \\\mathcal{B} &= \sqrt{\Gamma^2 - 4 \gamma_c \gamma_s \eta}.
\end{align}

From these equations we can obtain an overall enhancement in the sensitivity-bandwidth product
\begin{equation}\label{eq:gain}
(\mathcal{S}\times \mathcal{B})/(\mathcal{S}\times \mathcal{B})_{\gamma_s=0}=\frac{\gamma_c+\gamma_l}{\mathcal{B}}.
\end{equation}

For a given detection loss there exists an optimal squeezing factor that gives maximal enhancement, which differs from the threshold value where the maximum squeezing is achieved. 
This can be understood as follows.
The maximal detectable squeezing value is is bounded by the amount of optical loss.
The loss of squeezing can be seen as mixing with vacuum \cite{Grynberg2010}, therefore, above a certain value the increase in squeezing is not detectable any more, see Eq.\,\eqref{eq:noise_tf}. 
However, the signal deamplification is independent of the detection loss, and has a weaker dependence on the internal loss, see Eq.\,\eqref{eq:signal_tf}. 
Therefore increasing the internal gain above a certain level leads to a larger detected deamplification in the signal than suppression in the shot noise level.

We experimentally test the presence of an enhancement in the sensitivity-bandwidth product compared to the standard limit, and show the influence of the detection loss on it.

\noindent \textit{Experiment.---}
 \begin{figure}
	\includegraphics{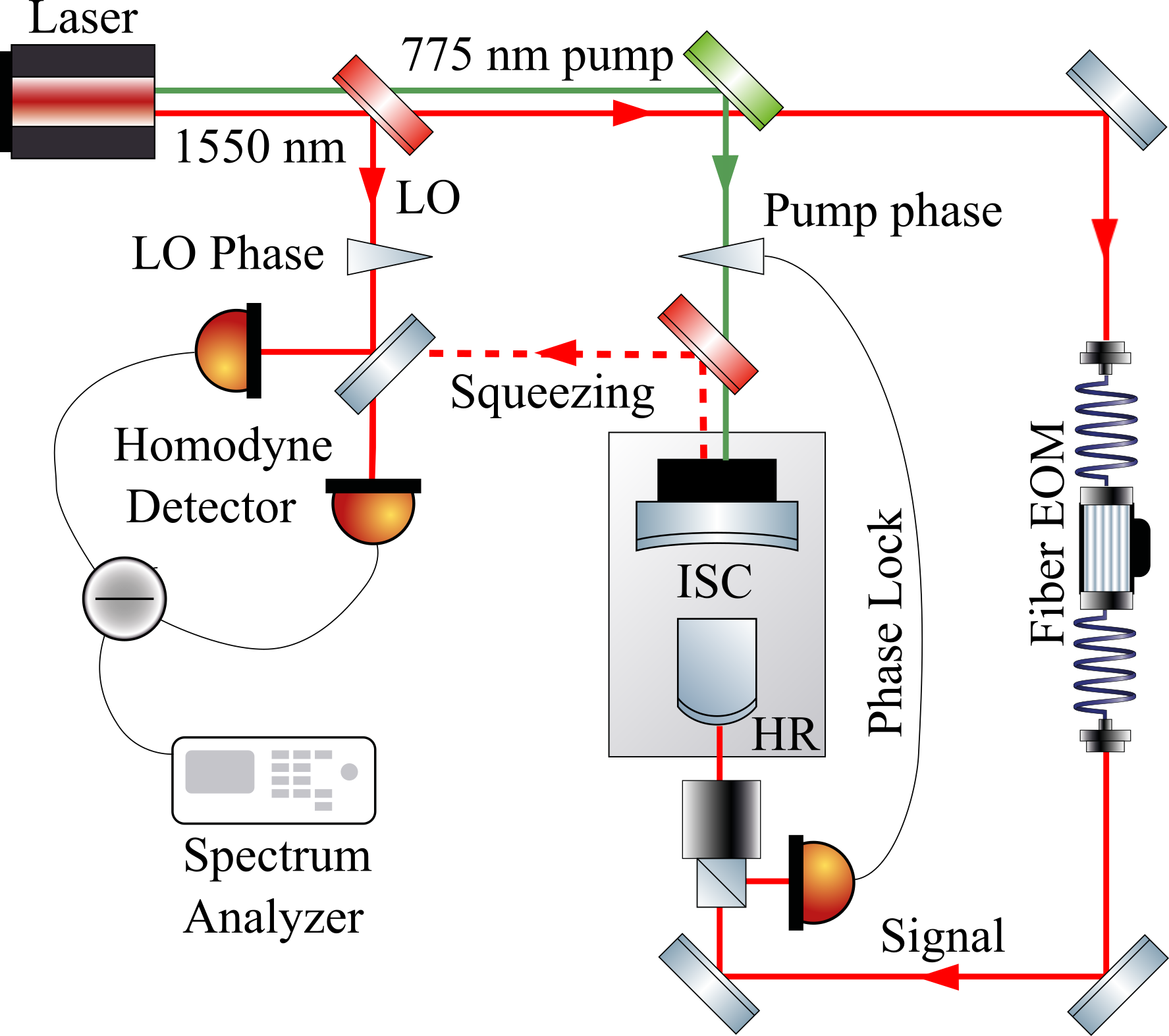}
	\onlinecap\caption{Experimental setup --- The internal squeezing cavity (ISC) is resonant for both the fundamental wavelength 1550\,nm and the second harmonic wavelength 775\,nm. Through the highly reflective back mirror a beam at 1550\,nm is injected carrying a phase modulation signal between 5.5 to 151\,MHz and a sideband at 54\,MHz for Pound-Drever-Hall (PDH) cavity length stabilization. The output signal consisting of squeezed light and deamplified signal sideband is detected on a balanced homodyne detector using 2.8\,mW local oscillator (LO) power, with an overall detection efficiency of $\sim85$\%. The phase of the local oscillator is actively stabilized to the phase quadrature, and the phase of the pump is stabilized to produce squeezing in the phase quadrature.}\label{fig:setup}
 \end{figure}
In our proof-of-principle experiment the signal is generated by injection of a phase modulated field from the back of the Fabry-Perot cavity with optical parametric amplifier inside.
In terms of signal detection and observation of the internal squeezing effect this approach can be viewed as one-to-one analogy to a detector with a movable end mirror sensitive to the external force.
The advantage of our approach is that it allows signal generation in a broad frequency band, which is necessary to observe the change in the detection bandwidth. 
 
The experimental setup, shown in Fig.\,\ref{fig:setup} consists of a second harmonic generation cavity (SHG), producing 775\,nm light for optical-parametric amplification of the longitudinal resonance at 1550\,nm of our internal squeezing cavity, here simply called ``ISC''. 
The cavity has an optical length of $L=2.77$\,cm, an optical linewidth of $ \gamma_c\sim 2 \pi \times 54$\,MHz, and contains a periodically poled KTP (PPKTP) crystal  \cite{Mehmet2011a}.
A control field at 1550\,nm with a phase-modulation signal imprinted on it is injected from the highly reflective back side of the ISC. 
The signal is produced by the broadband fiber electro-optical modulator (EOM). 
The cavity length is stabilized via the Pound-Drever-Hall (PDH) locking technique \cite{Drever1983, Black2001}. 
The ISC has two locking modes - with and without the pump light. When the measurements with squeezing are taken, the cavity length is stabilized with 775\,nm light, while the 1550\,nm control field is used to stabilize the squeezing angle on the phase quadrature. 
When the measurements without squeezing are taken, the 775\,nm pump is off, and the cavity length is stabilized with the 1550\,nm control field. 
The signal with or without squeezing is detected with a high-efficiency broadband homodyne detector with a bandwidth of $\sim$\,800\,MHz and dark noise clearance of $\sim$\,13\,dB in the frequency range of interest from 10 to 200\,MHz. 

We create a phase modulation signal at different frequencies. 
At each frequency we detect the signal together with the noise on the homodyne detector, in two regimes: with the optical parametrical amplification being on and off.
This allows us to observe how the signal gets deamplified, and noise --- squeezed.
From the squeezing spectrum we estimate the experimental parameters: squeezing factor $q$, transmissivity of the coupling mirror $t_{\rm c}$, internal loss $l^2$ and detection efficiency $\eta$. 
The fitted upper bound on the internal loss of $l^2\leq2300$\,ppm, which results in in the roundtrip loss bandwidth of $\gamma_l\leq 2\pi\times743\,kHz \ll \gamma_c$, is consistent with the previously measured absorption of a PPKTP crystal \cite{Steinlechner2013}, and the manufacturer specified transmissivity of the back mirror ($t^2_{\rm b} = 0.05\%$@1550\,nm)  and bound on the anti-reflective coating of the crystal ($r^2< 0.1\%$). 
The coupling mirror transmissivity of $t^2_{\rm c} =15\%$@1550\,nm is confirmed by an independent measurement of the cavity finesse. 
The detection loss estimation is also bounded within 1\% of the estimated value by comparing squeezing and anti-squeezing spectra \cite{Mehmet2011a,Vahlbruch2016}. 
We use these estimated parameters to calculate the expected theoretical spectrum of the signal deamplification and compare it with the measured values.

\begin{figure}
	\includegraphics{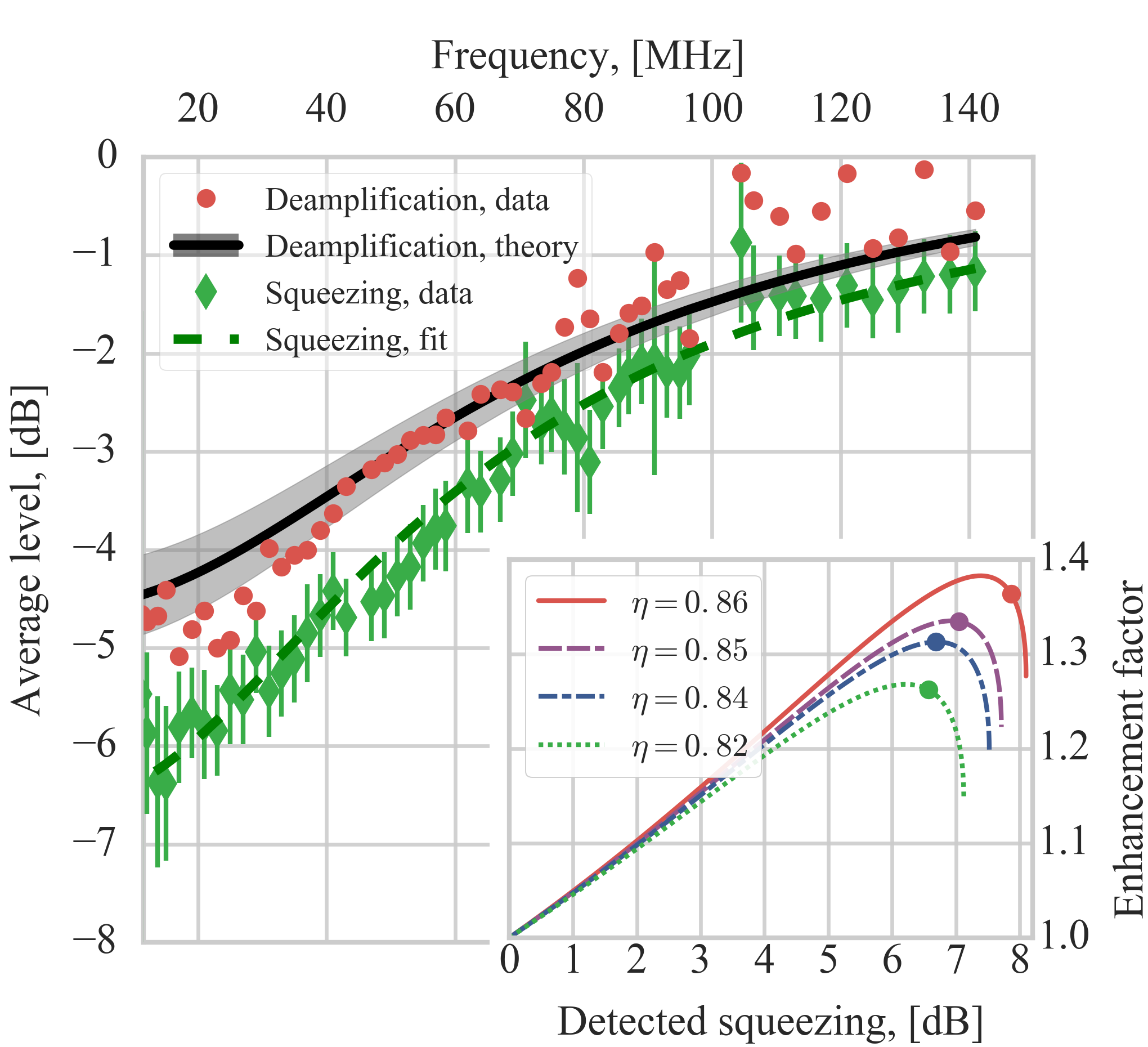}\onlinecap\caption{Beating the standard sensitivity-bandwidth limit with internal squeezing. In the main plot we demonstrate the increase in the signal-to-noise ratio for a total quantum efficiency of 0.82: squeezing data (green rhomb) with fit (green dashed line), from which the squeezing factor is estimated; the signal deamplification is represented by the red dots.
	It is compared to the results of the theoretical modelling (black solid line) with parameters obtained from the squeezing measurement, where the grey area represents the confidence interval based on the estimation error. The subplot shows four experimentally achieved enhancement factors (26\%, 31\%, 33\%, 36\%) representing four different overall quantum efficiencies, together with theoretical curves versus detected squeeze factor.    Two effects are demonstrated: the dependence on the detection efficiency $\eta$ (different curves) and the existence of the optimal squeezing for each set of parameters. The maximal enhancement in the sensitivity-bandwidth product obtained in the experiment is $36\%$ (red solid curve). The data on the main plot corresponds to the green (dotted) curve with $82\,\%$ detection efficiency and $26\%$ enhancement. }\label{fig:result}
 \end{figure}

Fig.\,\ref{fig:result} compares noise squeezing with signal deamplification; the difference between the two data sets directly demonstrates the increase in the SNR, corresponding to an enhancement of 26\% in the sensitivity-bandwidth product beyond the standard limit.
We find the theory to be in good agreement with the experimental data, lying within the confidence interval obtained from the parameter estimation error.
We ascribe the observed discrepancies to the electronic resonances in the homodyne circuitry and wires that are not taken into account in the theoretical analysis.
Higher enhancement factors are observed with a second homodyne detector, which has less loss (but also stronger electronic resonances).
The dots in subplot in Fig.\,\ref{fig:result} shows four experimentally achieved enhancement factors (26\%, 31\%, 33\%, 36\%) representing four different overall quantum efficiencies. 

\noindent
\textit{Summary and outlook.---}
In summary, we provide a unified view of three different nonclassical concepts for improving the quantum measurement noise limited sensitivity of cavity-enhanced laser interferometers. 
All of them can be seen as concepts of beating the standard sensitivity-bandwidth limit. 
Two of these concepts: ``white-light cavity'' and ``external squeezing'', have been investigated intensively in recent years for the improvement of gravitational-wave detectors \cite{Miao2105, Qin2015, Schnabel2016}.
In this work, the third concept, ``internal squeezing'', is investigated, theoretically as well as experimentally, and also in view of improving gravitational-wave detectors. 
We presented the first experimental demonstration of beating the standard sensitivity-bandwidth limit with internal squeezing. 

We note that all three concepts can in principle be combined to maximise the overall improvement. 
The most mature concept is external squeezing, as it is already implemented in the gravitational-wave detector GEO\,600 \cite{Abadie2011}. 
Since it avoids any deamplification of the signal and squeezes shot noise in a broadband way, it is more efficient than internal squeezing.
Interestingly, its sensitivity to intra-cavity loss is higher than that of internal squeezing. 
This can be understood in the limiting case when the cavity round trip loss equals the coupler's transmissivity. 
In this case, the cavity is impedance matched for external squeezing, and no squeezing gets reflected off the coupling mirror.
On contrary, in the internal squeezing case only half of the squeezing produced inside the cavity is lost. The other half is coupled out through the mirror, resulting in a maximal measurable squeeze factor of 3\,dB.
Based on our work, we thus propose to combine external and internal squeezing to improve the sensitivities of gravitational-wave detectors to values that are not possible with external squeezing alone.

\noindent\textit{Acknowledgements.---}
We thank Sacha Kocsis for comments that greatly improved the manuscript, Paolo Piergentili for assistance with the electronics' design, the members of MQM discussion group and LSC QNWG working group for many fruitful discussions.
Research of M.K. was supported by the Marie Curie Initial Training Network cQOM; research of M.K., L.K and R.S. is supported by the European Research Council (ERC) project ``MassQ'' (Grant No. 339897); research of S.A. was supported by the IMPRS on Gravitational-Wave Astronomy; research of H.M. is supported by UK STFC Ernest Rutherford Fellowship; research of R.S. is supported by Deutsche Forschungsgemeinschaft (Grant No. SCHN 757-6).

\appendix
\section{Supplementary material for ``Beating the standard sensitivity-bandwidth limit of cavity-enhanced interferometers with internal squeezed-light generation"}

   \subsection{Derivation of the noise spectrum and signal transfer function}
           \begin{figure}[t]
	\includegraphics{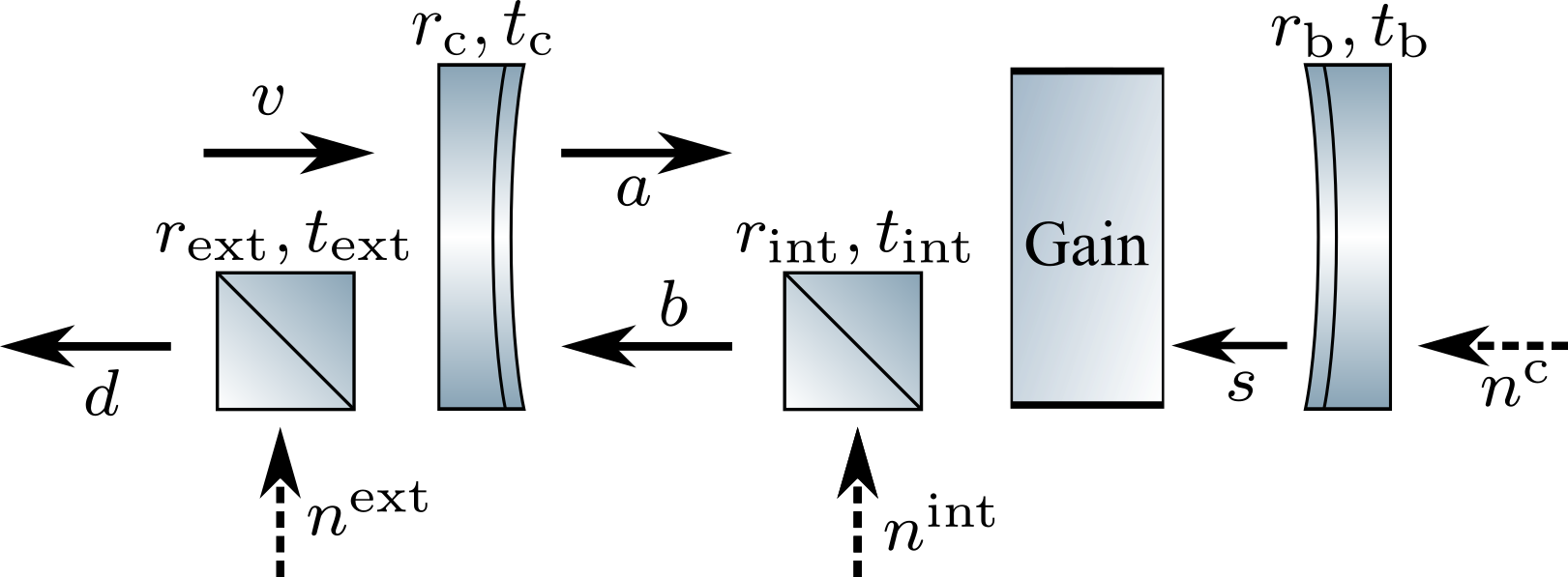}\caption{The schematic representation of our interferometric system. The signal $s = 2 i k_p x(\Omega)$ originates from the displacement $x(\Omega)$ of the back mirror which is assumed to be of infinite mass. Two sources of loss are assumed. Internal loss and detection loss, and both are modelled with a beam splitter. The optical parametrical amplification process creates a gain in one of the quadratures, and an attenuation in the orthogonal one.}\label{fig:sup_simple}
 \end{figure}
 In this section we derive the theoretical model for our interferometric system.
It corresponds to a Fabry-Perot cavity with nonlinear media inside, that parametrically amplifies one quadrature of the light (see Fig.\,\ref{fig:sup_simple}). 
  Using the perturbation theory, we decompose the light field into a steady-state amplitude with amplitude $A_0$ and laser carrier frequency $\omega_0$ and a slowly varying noise amplitude $a(t)$ (see details in \cite{Danilishin2012}):

  \begin{align}
  &A(t) = \sqrt{\frac{2\pi \hbar \omega_0}{\mathcal{A} c}} \left[ A_0 e^{-i\omega_0 t} + a(t) e^{-i\omega_0 t} \right] + {\rm h.c.}\\
  &a(t) = \int_{-\infty}^{\infty} a(\omega_0 + \Omega)e^{-i\Omega t} \frac{d \Omega}{2 \pi},
  \end{align}
  where $\mathcal{A}$ is the laser beam cross-section area, $\hbar$ is the reduced Plank constant. Note that we omit the hats on the operator for brevity, although all the fields are quantised.  Below we consider only the noise fields in the frequency domain. Furthermore, we define the two-photon amplitude and phase quadratures at a sideband frequency $\Omega$ correspondingly as
  \begin{align}
  a_x(\Omega) &= \frac{a(\omega_0 + \Omega) + a^\dag(\omega_0 - \Omega)}{\sqrt{2}},\\
  a_y(\Omega) &= \frac{a(\omega_0 + \Omega) - a^\dag(\omega_0 - \Omega)}{i\sqrt{2}}.
  \end{align}

  These operators obey the commutation relation
  \begin{align}
  &[a_x(\Omega), a_x(\Omega')] = [a_y(\Omega), a_y(\Omega')] = 0,\\
   &[a_x(\Omega), a_y(\Omega')] = [a_x(\Omega), a_y(\Omega')] = 2 \pi i \delta(\Omega + \Omega').
  \end{align}

  Using these two-photon quadratures, we can apply the input-output formalism \cite{Caves1985a, Schumaker1985a} and find the steady-state fields in the system.

 The signal we consider is a phase modulation on the light field induced by motion of the mirror with infinite mass caused by an external force.
 This modulation adds a phase shift on the light reflected off the movable mirror: $E_{\rm refl} = E_{\rm in}e^{2 i k x(\Omega)}\approx E_{\rm in}(1 + 2 i k_p x(\Omega))$, where $k_p$ is the light's wave vector, $E_{\rm refl, in}$ are the amplitudes of the reflected and incident light fields, and $x(\Omega)$ is a small mirror displacement.
 The signal appears only in the equations for the phase quadrature of the light field.
 We model the optical loss by a beamsplitter reflecting some part of the light fields to the environment and mixing in some vacuum from the environment. 
The optical parametrical amplification process is not included in the model.
We treat crystal as a gain medium that linearly amplifies with gain $e^{q}$ a certain quadrature (amplitude in our case) and deamplifies the orthogonal one. 
We call $q$ a squeezing factor in a single pass through the crystal.

The system of input-output equations for the amplitude (denoted by $x$) and phase (denoted by $y$) quadratures reads
\begin{widetext}
\begin{align}
&\begin{cases}
a_x (\Omega) = t_{\rm c} v_x (\Omega)+ r_{\rm c} b_x (\Omega)\\
b_x(\Omega) = a_x (\Omega) t_{\rm int} r_{\rm b}   e^{2 i \Omega \tau} e^{2 q}  + n^{\rm c}_x (\Omega) t_{\rm int} t_{\rm b}   e^{i\Omega \tau} e^{q} +  r_{\rm int}  n^{\rm int}_x (\Omega) \\
d_x  (\Omega)= t_{\rm det} \left(-r_{\rm c}  v_x (\Omega) + t_{\rm c}  b_x (\Omega)\right) + r_{\rm det} n^{\rm ext}_x (\Omega)
\end{cases}\\
&\begin{cases}
a_y (\Omega) = t_{\rm c} v_y (\Omega)+ r_{\rm c} b_y (\Omega)\\
b_y(\Omega) = a_y (\Omega) t_{\rm int} r_{\rm b} e^{2 i \Omega \tau } e^{-2 q}  + 2 i k_p E x(\Omega) t_{\rm int}  e^{i\Omega \tau} e^{-q} + n^{\rm c}_y (\Omega)  t_{\rm int} t_{\rm b}  e^{i\Omega\tau} e^{-q} + r_{\rm int} n^{\rm int}_y (\Omega) \\
d_y  (\Omega)= t_{\rm det} \left(-r_{\rm c} v_y (\Omega) + t_{\rm c} b_y (\Omega)\right) + r_{\rm det} n^{\rm ext}_y (\Omega).
\end{cases}
\end{align}
\end{widetext}
Here $\tau = L/c$ is the round trip propagation time, with $L$ being the length of the cavity; $c$ is the speed of light, $q$ - single pass squeeze factor, $x(\Omega)$ - mirror displacement induced by a signal, $k_p$ - wave vector of the carrier field, $E$ is the mean amplitude of the light field inside the detector, $t_{\rm c,b}, r_{\rm c,b}$ - amplitude transmissivity and reflectivity of coupling and back mirrors, such that $r_{\rm c,b}^2 + t_{\rm c,b}^2 = 1$, $r_{\rm det}$ - detection loss, $r_{\rm int}$ - intra-cavity loss without the coupling and the back mirrors; $r^2_{\rm det, int}+t^2_{\rm det, int}=1$.

This set of equations can be solved for the detected fields $d_{x,y}$:
\begin{widetext}
\begin{align}
d_x (\Omega) &= \frac{t_{\rm det}}{e^{-2q}-e^{2i\Omega \tau}r_{\rm c}r_{\rm b} t_{\rm int}} \left(v_x(\Omega)\left(-r_{\rm c} e^{-2q} + r_{\rm b} r_{\rm int}^2 e^{2i\Omega \tau} \right)+ n^{\rm c}_x(\Omega) t_{\rm c} t_{\rm b} t_{\rm int} e^{-q} e^{i\Omega\tau} + \right. \nonumber \\ 
&\left. + n^{\rm int}_x(\Omega) t_{\rm c} r_{\rm int} e^{-2q})\right) + r_{\rm det} n^{\rm ext}_x (\Omega),\\
d_y (\Omega)&= \frac{t_{\rm det}}{e^{2q}-e^{2i\Omega \tau}r_{\rm c}r_{\rm b} t_{\rm int}} \left(2 i k_p E x(\Omega) t_{\rm c} t_{\rm int} e^q e^{i\Omega \tau} +  v_y(\Omega)\left(-r_{\rm c} e^{2q} + r_{\rm b} t_{\rm int} e^{2i\Omega \tau} \right)+ n^{\rm c}_y(\Omega) t_{\rm c} t_{\rm b} t_{\rm int} e^q e^{i\Omega\tau} + \right. \nonumber \\ 
&\left. + n^{\rm int}_y(\Omega) t_{\rm c} r_{\rm int} e^{2q})\right) + r_{\rm det} n^{\rm ext}_y (\Omega).
\end{align}
\end{widetext}

The spectrum of the noise $a(\Omega)$ is defined:
\begin{equation}
S_{a} (\Omega) \delta(\Omega - \Omega') = \frac{1}{2}\left<a(\Omega)a(\Omega') + a(\Omega')a(\Omega)\right>.
\end{equation}
Then, assuming that all noises in the system are uncorrelated, we find the spectral density of the detected noise
\begin{equation}\label{eq:noisefull}
S_n(\Omega) = 1 - \frac{t_{\rm c}^2 t_{\rm det}^2t_{\rm int}^2(1-e^{-2q})(1 + e^{-2q} r_{\rm b}^2)}{1 + r_{\rm c}^2r_{\rm b}^2 t_{\rm int}^2 e^{-4q} -2 r_{\rm c} r_{\rm b} t_{\rm int} e^{-2 q} \cos 2\Omega\tau}.
\end{equation}

The transfer function of the signal $x(\Omega)$ through the optical cavity to the detector:
\begin{equation}
T(\Omega) = 2i k_p E \frac{t_{\rm c} t_{\rm det}t_{\rm int} e^q e^{i\Omega \tau}}{e^{2q}-e^{2i\Omega \tau}r_{\rm c}r_{\rm b} t_{\rm int}},
\end{equation}
and it's spectral shape:
\begin{equation} \label{eq:sigfull}
|T(\Omega)|^2=\frac{8 \pi P_c}{\hbar \lambda L} \frac{e^{-2q}t_{\rm c}^2 t_{\rm det}^2t_{\rm int}^2}{1 + r_{\rm c}^2r_{\rm b}^2 t_{\rm int}^2 e^{-4q} - 2 r_{\rm c}r_{\rm b} t_{\rm int} e^{-2 q} \cos 2\Omega\tau},
\end{equation}
where $P_c = \hbar k_p c |E|^2$ is the light power inside the cavity, and $\lambda$ is the carrier wavelength.

\subsection{Approximate description}
In this section we simplify the expressions for the spectral density \eqref{eq:noisefull} and signal transfer function \eqref{eq:sigfull} by making several assumptions. 
The amplitude transmissivities of the coupling and the back mirrors, as well as the internal loss, are much smaller than unity, and we can approximate correspondingly $r_{\rm c, b}\approx 1-t_{\rm c, b}^2/2$ and $t_{\rm int}\approx 1-r_{\rm int}^2/2$; the squeezing factor $q$ is much smaller than unity, so we can approximate $e^{q}\approx1+q$; the frequency of interest is much smaller than the FSR of the cavity $\Omega \ll 1/2\tau$, which enables us to make a Taylor expansion: $\cos \Omega \tau \approx 1- \Omega^2 \tau^2/2$. 
With these assumptions we introduce the variables
\begin{align}
\gamma_{\rm c} = \frac{c t_{\rm c} ^2}{4 L}, \quad \gamma_s = \frac{q c}{L}, \quad \gamma_l = \frac{c l^2}{4 L},\\
 l^2=r^2_{\rm int} + t^2_{\rm b}, \quad \Gamma = \gamma_c + \gamma_s + \gamma_l.
\end{align}
Equations \eqref{eq:noisefull}, \eqref{eq:sigfull} then simplify to
\begin{align}
|T(\Omega)|^2 &\approx \frac{8 \pi P_c}{\hbar \lambda L}\frac{\gamma_{\rm c}\eta}{\Gamma^2 + \Omega^2},\\
S_n(\Omega) &\approx 1 - \frac{4 \gamma_{\rm c} \gamma_s}{\Gamma^2 + \Omega^2} \eta,
\end{align}
where $\eta =1-t_{\rm det}^2$ is the detection efficiency.
Then the signal-to-noise ratio reads
\begin{equation}
\frac{|T(\Omega)|^2}{S_n(\Omega)} =  \frac{8 \pi P_c}{\hbar \lambda L} \frac{\gamma_{\rm c}\eta}{\Gamma^2 - 4 \gamma_{\rm c} \gamma_s\eta + \Omega^2},
\end{equation}
with corresponding bandwidth
\begin{equation}
\mathcal{B} = \sqrt{\Gamma^2 - 4 \gamma_{\rm c} \gamma_s\eta}.
\end{equation}
We define the integrated sensitivity, which connects to the sensitivity-bandwidth product
\begin{equation}\label{eq:int}
\rho = \int\nolimits_0^{\omega_{\rm FSR}} \frac{|T(\Omega)|^2}{S_{n}(\Omega)} d\Omega = \mathcal{S}\times \mathcal{B},\end{equation}
where $\omega_{\rm FSR}$ is a free spectral range of the cavity, and $\mathcal{S} = |T(0)|^2/S_{n}(0)$ is the peak sensitivity.
The enhancement in the sensitivity-bandwidth product is given by
\begin{equation}
G =\rho/\rho_{q=0} = \frac{\gamma_c + \gamma_l}{\mathcal{B}}.
\end{equation}

As we can see from the equations, the main source of reduction of the desired effect is the detection loss. It includes both optical loss in the path and quantum efficiency of the detector. 
\begin{figure*}
\begin{minipage}{0.48\textwidth}
\includegraphics{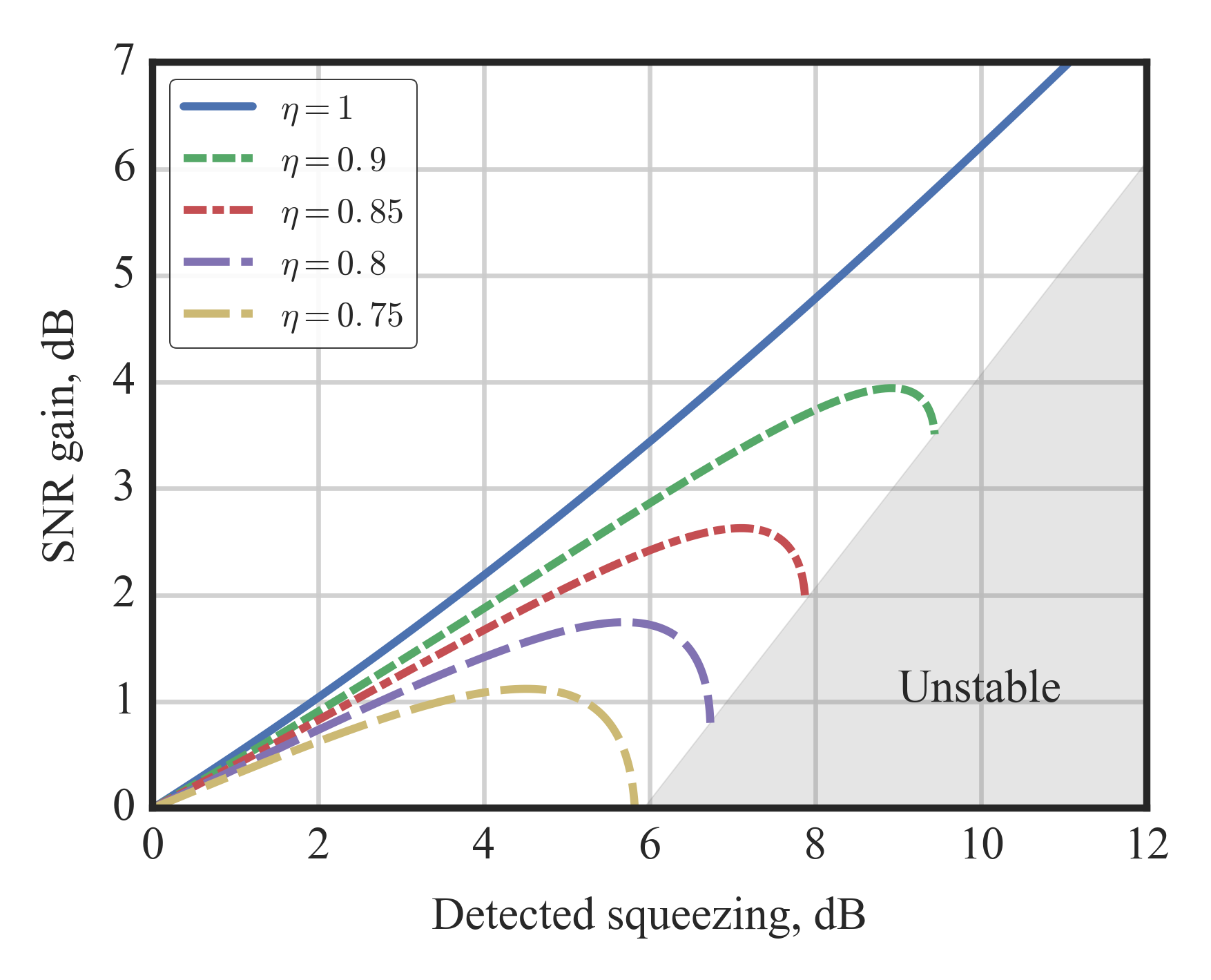}
\end{minipage}
\begin{minipage}{0.48\textwidth}
\includegraphics{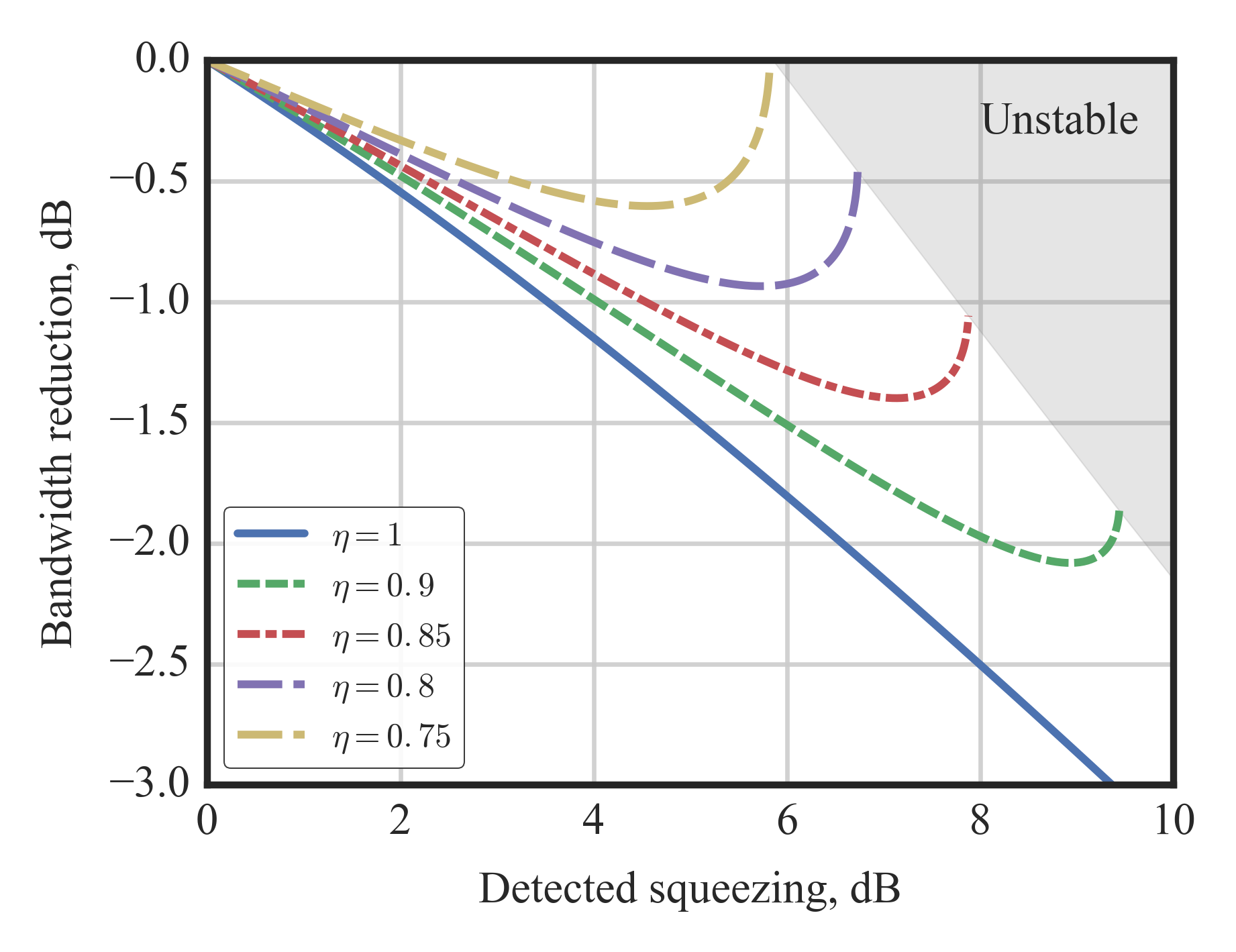}
\end{minipage}
\onlinecap
\caption{Dependence of the gain in the signal-to-noise ratio (left) and the reduction in the bandwidth (right) depending on the detected squeeze factor. Different plots represent the influence of the detection loss on the enhancement. The existence of an optimal squeezing is demonstrated. The shaded region represents the parametric gains for which the intra-cavity field becomes unstable.}\label{fig:sup_theory}
\end{figure*}
We can see in Fig.\,\ref{fig:sup_theory}  that there exists an optimal squeeze factor, at which the enhancement is maximal.

The treatment presented in this section is useful for understanding the concept and main properties of the internal squeezing, but in real cases some of the assumptions made here might be not valid.
Thus one has to calculate the integrated sensitivity \eqref{eq:int} directly from Eqs. \eqref{eq:noisefull}, \eqref{eq:sigfull}.

\subsection{Calculation of the optical parametric oscillation threshold}
In this section we derive the threshold value for the squeeze factor.
The squeezing value cannot be arbitrary large inside the cavity, as at some pump power it will become unstable and initiate lasing. We find the stability criterion from the equation for the amplitude quadrature inside the cavity:
\begin{equation}
b_x = \frac{v_x(\Omega) r_{\rm b} t_{\rm c} t_{\rm int} e^{2q} e^{2 i \Omega \tau} + n^{\rm c}_x (\Omega) t_{\rm b}e^q e^{i \Omega \tau} + n^{\rm int}_x(\Omega) r_{\rm int}}{1 - r_{\rm c} r_{\rm b} t_{\rm int}e^{2 q}e^{2i\Omega\tau}}.
\end{equation}
The threshold value represents the condition, at which the gain becomes larger than the overall loss through the coupler transmission and additional optical round trip loss. This condition is defined by setting the denominator equal to zero, and is reached when
\begin{equation}
e^{2q} = \frac{1}{ r_{\rm c} r_{\rm b} t_{\rm int} }.
\end{equation}

\subsection{Maximal squeeze factor inside the cavity}
Here we compare the maximally achievable squeezing inside the cavity in frequency and time domain, bringing the results of Ref. \cite{Milburn1981,Collett1984} in accord with our notations.

Inside the cavity the squeezing spectrum of the phase quadrature is (in the assumption of frequency being much smaller than the cavity FSR):
\begin{equation}\label{eq:sqz_limit}
S^{\rm in}(\Omega) = \frac{r_{\rm c}^2 r_{\rm int}^2 + t_{\rm c}^2 + r_{\rm c}^2 t_{\rm b}^2 t_{\rm int}^2 e^{-2q} + t_{\rm c}^2 r_{\rm b}^2 r_{\rm int}^2 t_{\rm int}^2 e^{-4q}}{(1-r_{\rm c}r_{\rm b}t_{\rm int}^2 e^{-2q})^2 + 4 e^{-2q} r_{\rm c}r_{\rm b}t_{\rm int}^2 \Omega^2 \tau^2}.
\end{equation}

At the threshold value in the limiting case \mbox{$r_{\rm f,b} \rightarrow 1, r_{\rm int} \rightarrow 0$} the amount of squeezing approaches
\begin{equation}
S^{\rm in}(0)/S^{\rm in}(0)_{q=0} \rightarrow \frac{1}{4},
\end{equation}
which is what we call the 6 dB squeezing limit. On the other hand, sometimes in the literature one can find 3 dB as the intra-cavity limit for squeezing. This limit refers to the maximal reduction in the noise variance of the cavity mode. Indeed, by integrating the spectrum \eqref{eq:sqz_limit} over the full frequency range and applying the same limits one finds the value of $1/2$ as a limit.
\twocolumngrid
\bibliography{internal_squeezing}

\end{document}